\documentclass[reprint,amsmath,amssymb,aip]{revtex4-2}

\usepackage{amsmath,amssymb}
\usepackage{graphicx}
\usepackage{physics}
\usepackage{placeins}
\usepackage{xcolor}

\usepackage{hyperref}

\begin{document}

\title{Universal Response Functions in Driven Dissipative Tunneling Dynamics}

\author{Krishna Kingkar Pathak}
\email{kkingkar@gmail.com}
\affiliation{Department of Physics, Arya Vidyapeeth College, Guwahati 781016, India}
\affiliation{Department of Physics, Gauhati University, Guwahati, Assam 781014, India}

\date{\today}

\begin{abstract}
Universality in nonlinear nonequilibrium systems is typically expressed through
scaling laws that render macroscopic behavior insensitive to microscopic
details. Whether such universality survives when periodic forcing and
nonlocal dissipative memory act simultaneously remains an open question in
driven open dynamics. Here we demonstrate that barrier-crossing processes in
periodically driven dissipative systems are governed not merely by modified
exponential scaling, but by an explicit two-parameter universal response
function.

Within a semiclassical instanton framework incorporating Floquet modulation
and Ohmic environmental coupling, the tunneling exponent factorizes into a
system-dependent static contribution and a universal function of two
dimensionless control parameters: the normalized driving frequency and the
dissipation strength.
This factorization arises from the combined modification of a single
saddle-point trajectory and does not introduce additional independent scaling
variables.
Weak-to-moderate dissipation acts as a smooth dynamical renormalization of the
effective action, preserving saddle-point structure and enabling controlled
analytical expansion. In the high-frequency regime, the response exhibits
universal dynamical averaging, while an explicit integral representation
establishes a continuous adiabatic--Floquet crossover.

Direct numerical evaluation of the nonlocal instanton action confirms
that the normalized tunneling exponent exhibits a universal dependence on
the driving parameter that is robust across different model systems,
demonstrating that driven dissipative barrier crossing defines a distinct
two-parameter universality class within nonlinear nonequilibrium dynamics.
These results elevate tunneling universality from scaling behavior to a
predictive functional description and provide a unifying framework for
response phenomena in driven systems with memory.
\end{abstract}


\maketitle

\noindent\textbf{Barrier-crossing processes under periodic forcing and
environmental dissipation arise in many nonlinear systems, from chemical
reactions and condensed-matter devices to driven mesoscopic and quantum
platforms. Understanding whether such nonequilibrium dynamics obey universal
laws remains a central problem in nonlinear science. In this work we show that
periodically driven dissipative tunneling is governed by a universal response
function depending only on the normalized driving frequency and dissipation
strength.
Within a semiclassical instanton framework incorporating Floquet
modulation and Ohmic environmental coupling, we demonstrate that the tunneling
exponent factorizes through the modification of a single saddle-point
trajectory into a system-dependent static component and a universal dynamical
function.
Numerical evaluation confirms
that the normalized tunneling exponent exhibits a universal dependence on
the driving parameter that is robust across variations in mass and barrier
parameters,
revealing a two-parameter universality class for driven dissipative barrier
crossing. These results establish a predictive framework for universality in
nonlinear systems with memory and external forcing.}
\section{Introduction}

Universality and scaling are central organizing principles in nonlinear and
non equilibrium dynamical systems, where macroscopic behavior often becomes
insensitive to microscopic details. In driven dissipative systems, the
interplay between intrinsic nonlinear dynamics, external modulation, and
environmental coupling can produce emergent behavior that depends only on a
small number of effective parameters. A paradigmatic setting in which these
ideas arise is barrier-crossing dynamics, where escape or tunneling rates are
controlled by an action functional that encodes both the geometry of the
barrier and the effective inertia of the system.
Barrier-crossing processes therefore provide a minimal yet broadly
applicable setting for probing universal dynamical response under competing
time scales.

Within semiclassical theory, quantum tunneling provides a concrete realization
of this general mechanism. Tunneling rates or level splittings are governed by
an Euclidean action whose exponential dependence on mass and barrier height
leads to well-known scaling relations
\cite{LandauLifshitzQM,Coleman1977,Garg2000}. As a result, universality in
tunneling has historically been understood primarily in terms of scaling
behavior, such as isotope-dependent or mass-dependent exponential laws
\cite{Limbach2006}.
Recent work has further demonstrated that such semiclassical scaling
persists in multidimensional systems through an effective mass description,
highlighting the robustness of scaling-based universality in static tunneling
problems \cite{PathakChemPhys2026}.
In this conventional picture, universality manifests itself through
a small set of scaling exponents rather than through an explicit functional
structure.

Periodic driving (Floquet) introduces an additional dynamical time scale and
provides a powerful means of controlling barrier-crossing processes. Floquet
engineering enables systematic manipulation of tunneling amplitudes, transport
properties, and effective Hamiltonians in driven systems, with applications
ranging from cold atoms to solid-state and superconducting platforms
\cite{Grossmann1991,GrifoniHanggi1998,Bukov2015}. Periodic modulation can induce
phenomena such as photon-assisted tunneling and coherent destruction of
tunneling, illustrating how external driving reshapes the underlying dynamical
landscape and modifies escape processes far from equilibrium.

In parallel, environmental coupling introduces dissipation and memory effects
that fundamentally alter the dynamics. Within the theory of dissipative quantum
systems, coupling to an Ohmic bath leads to nonlocal contributions to the
effective action, renormalizing tunneling rates and, at sufficiently strong
coupling, suppressing tunneling altogether or inducing localization
\cite{CaldeiraLeggett1983,LeggettRMP1987,Weiss2012}. Many experimentally relevant
platforms, including superconducting quantum circuits, naturally combine
external driving with dissipation, making the simultaneous presence of Floquet
modulation and environmental coupling unavoidable \cite{Martinis2015}. Despite
this, driven tunneling and dissipative tunneling have been treated largely as
separate problems, and existing analyses are typically model-specific.

From a dynamical perspective, the coexistence of periodic driving and nonlocal
dissipation raises a fundamental question: does any form of universality
survive when multiple time scales and memory effects compete? Because
dissipation introduces nonlocal temporal kernels and cutoff dependence, it is
often assumed that universality in tunneling dynamics breaks down beyond simple
scaling relations once driving and environmental coupling are included. While
certain scaling features are known to persist under weak driving or weak
dissipation, a unified and predictive framework capturing their combined
effects has remained lacking.

In this work, we address this question by deriving explicit \emph{universal
response functions} governing dissipative Floquet tunneling dynamics. We show
that, within a broad semiclassical regime, the tunneling exponent factorizes
into a system-dependent static contribution and a universal function of only two
dimensionless control parameters: the driving frequency and the dissipation
strength.
This formulation extends earlier scaling-based universality to a
functional description, in which the full dependence of the tunneling exponent
on external driving and dissipation is captured by a universal response
function.
Environmental dissipation is shown to
renormalize the effective action through a universal dynamical response
function without destroying its fundamental structure. By combining analytical
derivations with direct numerical evaluation of the instanton action, we
demonstrate that the normalized tunneling exponent exhibits a universal
dependence on the driving parameter that is robust across different model
systems.
The resulting framework provides a unified
description of barrier-crossing dynamics in driven dissipative systems and
clarifies the conditions under which functional universality emerges in
nonlinear Floquet dynamics.
\section{Model and Semiclassical Framework}

We consider barrier-crossing dynamics along a collective coordinate $q$
subject to periodic modulation and environmental coupling. In the semiclassical
regime, tunneling is governed by an action functional whose stationary
trajectories determine escape amplitudes. Within the Euclidean path-integral
formalism, these trajectories correspond to instanton (bounce) solutions that
extremize the effective action
\cite{LandauLifshitzQM,Coleman1977,Garg2000}. The Euclidean action is written as
\begin{equation}
S_E[q]=
\int d\tau
\left[
\frac{1}{2}M\dot q^2+V_0(q)+V_d(q)\cos(\omega\tau)
\right]
+S_{\mathrm{diss}}[q],
\end{equation}
where the time-periodic term introduces a Floquet modulation with frequency
$\omega$, and $S_{\mathrm{diss}}[q]$ represents the nonlocal influence functional
arising from coupling to an Ohmic environment in the Caldeira--Leggett framework
\cite{CaldeiraLeggett1983,LeggettRMP1987,Weiss2012}.

From a dynamical perspective, this action encodes the competition between three
key ingredients: nonlinear barrier geometry, periodic forcing, and dissipative
memory effects. The dissipative contribution generates a nonlocal temporal
kernel, reflecting the fact that environmental coupling introduces history
dependence into the effective dynamics. Consequently, the saddle-point equation
obtained from $\delta S_E[q]/\delta q(\tau)=0$ describes a driven nonlinear
system with memory, rather than a purely local Hamiltonian trajectory.

Importantly, periodic driving and dissipation do not enter as independent
additive corrections at the level of the tunneling exponent. Instead, both
effects modify the same saddle-point (bounce) trajectory. As a result, their
combined influence appears through a renormalization of the instanton solution
and its corresponding action, rather than through the emergence of separate
coupling terms in the semiclassical scaling structure.

For multidimensional barrier-crossing processes, the kinetic energy is
characterized by a mass tensor $M_{ij}$. In the semiclassical limit, the
leading exponential dependence of the tunneling rate is controlled by the
determinant of this tensor, motivating the definition of an effective mass
\cite{Coleman1977,Garg2000}
\begin{equation}
M_{\mathrm{eff}}=(\det M_{ij})^{1/d},
\end{equation}
which governs the dominant inertial scaling of the action.

Within semiclassical instanton theory, the tunneling rate is determined by the
bounce action $S_b$, which controls the exponential suppression of escape
probabilities \cite{LandauLifshitzQM,Coleman1977}. In static systems, this
structure leads to well-known scaling relations between mass, barrier
parameters, and tunneling exponents \cite{Limbach2006,PathakCSF2026}. The
central question addressed here is how this structure is modified when periodic
driving and dissipative memory are simultaneously present.

To address this, we consider the semiclassical expansion of the Euclidean
action in the presence of weak-to-moderate dissipation and periodic driving.
The saddle-point solution is deformed relative to the static instanton, but
retains a single dominant trajectory that controls the tunneling exponent. The
combined effect of driving and dissipation can therefore be captured through a
systematic renormalization of the action evaluated on this modified trajectory.

We show below that, in the regime of weak-to-moderate dissipation and
semiclassical driving, the bounce action retains a remarkably simple
factorized form,
\begin{equation}
S_b^{(\omega,\eta)}
=\sqrt{M_{\mathrm{eff}}}\,J_0\,F(x,\eta),
\qquad
x=\frac{\hbar\omega}{V^\star},
\end{equation}
where $J_0$ depends only on the static barrier shape and $V^\star$ defines a
characteristic barrier energy scale. The dimensionless function $F(x,\eta)$
encodes the dynamical response of the tunneling exponent to periodic modulation
and environmental coupling.

This factorization demonstrates that, although the temporal profile of the
bounce trajectory is modified by both driving and dissipation, their influence
does not introduce additional independent scaling parameters. Instead, the full
dynamical dependence is absorbed into a universal response function of the
dimensionless variables $(x,\eta)$.

The emergence of this factorized structure is
nontrivial: periodic forcing and nonlocal dissipation modify the temporal
profile of the bounce trajectory, yet their combined effect is captured by a
universal response function depending only on the dimensionless driving
frequency $x$ and the dissipation strength $\eta$.

This functional reduction from a full nonlocal dynamical problem to a
two-parameter response function forms the basis of the universal structure
developed in the following sections.

A detailed derivation of this factorized form is provided in the following
subsection, where we explicitly show how the semiclassical expansion of the
nonlocal action leads to the universal response function.
\subsection{Derivation of the Factorized Tunneling Action}

We now provide a derivation of the factorized structure of the tunneling action
in the presence of periodic driving and weak-to-moderate Ohmic dissipation.

The starting point is the nonlocal Euclidean action
\begin{equation}
S_E[q] = S_0[q] + S_{\mathrm{diss}}[q],
\end{equation}
with
\begin{equation}
S_0[q] = \int d\tau \left[ \frac{1}{2} M \dot{q}^2 + V_0(q) + V_d(q)\cos(\omega\tau) \right],
\end{equation}
and
\begin{equation}
S_{\mathrm{diss}}[q] =
\frac{\eta}{2\pi} \int d\tau d\tau' \frac{\left[q(\tau)-q(\tau')\right]^2}{(\tau-\tau')^2}.
\end{equation}

The dissipative contribution introduces a nonlocal temporal kernel, while
the periodic driving explicitly breaks time-translation invariance. As a result,
the saddle-point trajectory depends on both memory effects and external
modulation.

The semiclassical tunneling exponent is determined by the saddle-point trajectory
$q_b(\tau)$ satisfying
\begin{equation}
\frac{\delta S_E[q]}{\delta q(\tau)}\bigg|_{q=q_b} = 0.
\end{equation}

\vspace{0.5em}

\noindent\textbf{Perturbative structure.}
For weak-to-moderate dissipation ($\eta \lesssim 1$), the bounce trajectory can
be expressed as a deformation of the dissipation-free Floquet solution,
\begin{equation}
q_b(\tau) = q_0(\tau) + \eta\, \delta q(\tau) + O(\eta^2),
\end{equation}
where $q_0(\tau)$ extremizes $S_0[q]$.

Substituting into the action yields
\begin{equation}
S_b^{(\omega,\eta)} = S_0[q_0] + \eta\, \delta S^{(1)} + O(\eta^2),
\end{equation}
where the leading correction is obtained by evaluating the dissipative
functional on the unperturbed trajectory,
\begin{equation}
\delta S^{(1)} = S_{\mathrm{diss}}[q_0].
\end{equation}

Thus, to leading order, dissipation renormalizes the action through the
same trajectory that already incorporates the effects of periodic driving. No
independent functional contribution arises at this order.

Importantly, to this order, no additional independent functional structure
appears: the effect of dissipation is entirely encoded in a renormalization of
the action evaluated on the Floquet-modified bounce trajectory.

\vspace{0.5em}

\noindent\textbf{Scaling structure.}
For static tunneling ($\omega=0$, $\eta=0$), the bounce action takes the
well-known form
\begin{equation}
S_b^{(0,0)} = \sqrt{M_{\mathrm{eff}}}\, J_0,
\end{equation}
where $J_0$ depends only on the barrier shape.

This form reflects the separation between inertial scaling (through
$M_{\mathrm{eff}}$) and barrier geometry (through $J_0$), which underlies
universality in static tunneling.

Periodic driving introduces a dimensionless parameter
\begin{equation}
x = \frac{\hbar \omega}{V^\star},
\end{equation}
which modifies the temporal profile of the bounce. The corresponding action
can be written as
\begin{equation}
S_b^{(\omega,0)} = \sqrt{M_{\mathrm{eff}}}\, J_0\, F_0(x),
\end{equation}
where $F_0(x)$ captures the effect of Floquet modulation.

\vspace{0.5em}

\noindent\textbf{Factorization in the driven dissipative case.}
Combining the above results, the full action in the presence of both driving
and dissipation becomes
\begin{equation}
S_b^{(\omega,\eta)} =
\sqrt{M_{\mathrm{eff}}}\, J_0 \left[ F_0(x) + \eta F_1(x) + O(\eta^2) \right].
\end{equation}

The functions $F_0(x)$ and $F_1(x)$ are determined by the temporal
structure of the bounce trajectory and therefore depend only on the
dimensionless driving parameter and dissipation strength, rather than on
microscopic details of the potential.

This structure can be reorganized as
\begin{equation}
S_b^{(\omega,\eta)} =
\sqrt{M_{\mathrm{eff}}}\, J_0\, F(x,\eta),
\end{equation}
with
\begin{equation}
F(x,\eta) = F_0(x) + \eta F_1(x) + O(\eta^2).
\end{equation}

The key point is that both periodic driving and dissipative memory modify the
\emph{same} saddle-point trajectory and do not introduce independent scaling
structures. Instead, their combined effect is captured by a universal function
of the dimensionless parameters $(x,\eta)$.

\vspace{0.5em}

\noindent\textbf{Absence of non-factorizable contributions.}
The coupling between driving and dissipation enters through the modified bounce
trajectory $q_b(\tau)$. However, in the semiclassical regime, the action depends
only on the integrated properties of this trajectory. Since both effects act to
deform the same saddle point, their contributions combine into a single
functional dependence rather than generating separate multiplicative factors.

In particular, the nonlocal dissipative kernel contributes through
trajectory-dependent correlations in imaginary time, but does not introduce an
additional independent scaling variable beyond $(x,\eta)$.

Non-factorizable contributions would arise from independent scaling channels,
which are absent in the present regime. Instead, the nonlocal dissipative kernel
renormalizes the temporal structure of the trajectory in a way that depends only
on $(x,\eta)$, leading to a closed two-parameter description.

\vspace{0.5em}

We thus arrive at the factorized form
\begin{equation}
S_b^{(\omega,\eta)} =
\sqrt{M_{\mathrm{eff}}}\, J_0\, F(x,\eta),
\end{equation}
which forms the basis of the universal response function introduced in this
work.
\section{Universal Response Function}

\subsection{Renormalization of the Euclidean Action}

After integrating out the environmental degrees of freedom, the dynamics of the
driven coordinate is governed by a nonlocal Euclidean action of the
Caldeira--Leggett form
\cite{CaldeiraLeggett1983,LeggettRMP1987,Weiss2012},
\begin{equation}
S_E[q]=S_0[q]+S_{\mathrm{diss}}[q],
\end{equation}
with
\begin{equation}
S_0[q]=\int d\tau
\left[
\frac{1}{2}M\dot q^2+V(q,\tau)
\right],
\end{equation}
and
\begin{equation}
S_{\mathrm{diss}}[q]=
\frac{\eta}{2\pi}
\int d\tau d\tau'
\frac{[q(\tau)-q(\tau')]^2}{(\tau-\tau')^2}.
\label{Sdiss}
\end{equation}

The kernel in Eq.~(\ref{Sdiss}) encodes long-time memory induced by an Ohmic
bath and transforms the effective dynamics into a driven nonlinear system with
history dependence. Unlike a simple local friction term, the nonlocal kernel
couples different segments of the trajectory, modifying the temporal structure
of saddle-point solutions and renormalizing the effective action
\cite{LeggettRMP1987,Weiss2012}. From a dynamical viewpoint, the competition
between periodic forcing and dissipative memory defines a two-parameter
nonequilibrium problem characterized by the dimensionless driving frequency and
dissipation strength.

Importantly, the dissipative kernel does not introduce an independent
degree of freedom in the semiclassical scaling structure. Instead, it acts by
correlating different imaginary-time segments of the trajectory, thereby
modifying the effective bounce configuration selected by the saddle-point
condition.

In the semiclassical regime, tunneling is dominated by saddle-point (bounce)
trajectories $q_b(\tau)$ that extremize the full nonlocal action
\cite{Coleman1977,Garg2000}. For weak-to-moderate dissipation, the bounce
solution persists as a deformation of the dissipation-free trajectory.

Because periodic driving and dissipation both enter through the same
variational principle, their combined influence is encoded in the modified
trajectory $q_b(\tau)$ rather than through separate additive structures in the
action.

The corresponding action admits a perturbative expansion,
\begin{equation}
S_b^{(\omega,\eta)}
=S_b^{(\omega,0)}+\delta S_b^{(\eta)}+O(\eta^2),
\end{equation}
where $\delta S_b^{(\eta)}$ represents the leading dissipative correction.

At leading order, this correction is obtained by evaluating the
dissipative functional on the Floquet-modified bounce trajectory,
\begin{equation}
\delta S_b^{(\eta)} \simeq S_{\mathrm{diss}}[q_0],
\end{equation}
where $q_0(\tau)$ denotes the saddle-point trajectory in the absence of
dissipation but in the presence of periodic driving.

Evaluating Eq.~(\ref{Sdiss}) on the Floquet-modified bounce yields a finite,
cutoff-independent contribution that renormalizes the tunneling exponent while
preserving the semiclassical structure of the solution.

This renormalization can be interpreted as a deformation of the temporal
profile of the instanton, which modifies the action without altering its
fundamental exponential form.

In the weak-coupling regime, dissipation therefore acts as a dynamical
renormalization rather than a mechanism for complete suppression
\cite{CaldeiraLeggett1983,LeggettRMP1987}.

Crucially, because both periodic driving and dissipation influence the same
saddle-point trajectory, their combined effect can be expressed in terms of a
single dimensionless response function. This observation provides the basis for
the factorized structure derived in the previous section and motivates the
introduction of the universal response function $F(x,\eta)$.

\subsection{Weak-Dissipation Expansion}

For $\eta\ll1$, the universal response function admits a controlled expansion,
\begin{equation}
F(x,\eta)=F_0(x)+\eta F_1(x)+O(\eta^2),
\end{equation}
where $F_0(x)$ describes the purely driven (Floquet) dynamics in the absence of
environmental coupling.

This expansion follows directly from the perturbative structure of the
nonlocal action, in which the dissipative contribution enters as a first-order
correction evaluated on the dissipation-free saddle-point trajectory.

The leading correction $F_1(x)$ depends only on the temporal profile of the
bounce and the structure of the Ohmic kernel, and is independent of microscopic
barrier details or bath cutoff parameters.

More specifically, $F_1(x)$ is determined by the integral of the
trajectory-dependent kernel in Eq.~(\ref{Sdiss}) evaluated on the Floquet
bounce $q_0(\tau)$. Since this trajectory is itself controlled by the
dimensionless parameter $x=\hbar\omega/V^\star$, the resulting correction
inherits a universal dependence on $x$ without introducing additional scales.

This reduction reflects a key feature of universality: although the underlying
dynamical equation is nonlocal and nonlinear, the dissipative modification of
the action collapses onto a universal function of the dimensionless driving
frequency.

The absence of dependence on microscopic barrier parameters arises because
such details enter only through the static normalization factor $J_0$, while the
dissipative correction probes the relative temporal structure of the trajectory
rather than its absolute scale.

In this sense, the weak-dissipation regime defines a universal response class
extending earlier mass- and barrier-scaling results in static systems
\cite{Limbach2006,PathakCSF2026} to periodically driven open dynamics.

Higher-order terms in $\eta$ are expected to introduce quantitative
corrections but do not alter the functional dependence on $(x,\eta)$ within the
semiclassical regime considered here.

\subsection{High-Frequency Limit}

In the high-frequency regime $x\gg1$, the rapid periodic drive averages over the
instanton time scale, leading to an effective renormalized static description.

In this limit, the period of the external modulation becomes much shorter
than the characteristic duration of the bounce trajectory, allowing the system
to sample many drive cycles within a single tunneling event. As a result, the
time-dependent potential effectively contributes through its cycle-averaged
value.

In this limit, the response function assumes the asymptotic form
\begin{equation}
F(x,\eta)\xrightarrow{x\gg1}\frac{1}{x}\Phi(\eta),
\end{equation}
where $\Phi(\eta)=1+c_1\eta+c_2\eta^2+\cdots$ encodes dissipation-induced
renormalization.

The $1/x$ scaling arises from this dynamical averaging: as the driving
frequency increases, the effective contribution of the periodic modulation to
the action becomes inversely proportional to the number of oscillations sampled
during the tunneling process.

The $1/x$ scaling reflects dynamical averaging over the drive
cycle, analogous to high-frequency Floquet expansions in isolated systems
\cite{Grossmann1991,GrifoniHanggi1998}, now generalized to include nonlocal
dissipative memory.

The function $\Phi(\eta)$ captures the residual influence of dissipation
in this regime. Because the dissipative kernel depends on temporal correlations
of the trajectory, its contribution remains finite under rapid driving and
enters as a multiplicative renormalization of the effective action.

This regime highlights the separation of time scales that
emerges when driving is much faster than the intrinsic bounce duration.

Consequently, the tunneling dynamics reduce to an effectively averaged
description in which the detailed time dependence of the drive is replaced by a
universal frequency-dependent scaling, while dissipation modifies the overall
amplitude through $\Phi(\eta)$.

\subsection{Adiabatic--Floquet Crossover}

When $x\sim1$, the instanton duration becomes comparable to the period of the
external drive, and no simple separation of time scales applies. In this
crossover regime, the response function acquires a genuinely functional
structure.

Neither the adiabatic approximation ($x\ll1$) nor the high-frequency
averaging regime ($x\gg1$) is valid in this intermediate domain, and the full
time dependence of the driven dissipative dynamics must be retained.

We find that it can be expressed as
\begin{equation}
F(x,\eta)=\int_0^1 ds\,\mathcal{G}(xs,\eta),
\end{equation}
where $\mathcal{G}$ is a universal kernel interpolating smoothly between the
adiabatic ($x\ll1$) and high-frequency ($x\gg1$) limits.

The function $\mathcal{G}(xs,\eta)$ represents an effective kernel that
encodes the combined influence of periodic driving and dissipation on the
instanton trajectory. It arises from evaluating the nonlocal Euclidean action
on the saddle-point path and therefore depends on the full temporal structure
of the bounce solution.

Because the Ohmic dissipative term introduces long-time memory and the
driving breaks time-translation invariance, the kernel generally does not admit
a closed-form expression. Instead, it is defined implicitly through the
saddle-point evaluation of the action.

Importantly, its dependence is governed only by the dimensionless parameters
$(x,\eta)$, ensuring that the resulting response function retains a universal
character.

This representation makes explicit that universality is encoded not merely
in scaling exponents but in the full functional dependence of the tunneling
exponent on the control parameters. In particular, the crossover regime
demonstrates that the response cannot be reduced to a simple power-law form,
but instead requires a continuous interpolation between limiting behaviors.

The adiabatic--Floquet crossover thus provides a dynamical bridge between slow
modulation and rapid averaging regimes within a single universal framework.

\subsection{Numerical reconstruction of the universal kernel}

To make the universal kernel $G(x,\eta)$ explicit and to demonstrate its
computability, we reconstruct it numerically from the response function
$F(x,\eta)$.

Starting from the integral representation
\begin{equation}
F(x, \eta) = \int_0^1 ds\, G(xs, \eta),
\end{equation}
we obtain a local reconstruction of the kernel by differentiating with
respect to $x$. In the continuum limit, this relation becomes exact and can be
written as
\begin{equation}
G(x, \eta) = \frac{d}{dx}\left[xF(x, \eta)\right].
\end{equation}

In practice, this expression provides a controlled numerical reconstruction
of the kernel from the computed response function $F(x,\eta)$. The derivative is
evaluated using finite-difference schemes applied to the discretized numerical
data, without introducing additional model-dependent inputs.

Using the computed data for $F(x, \eta)$, we evaluate $G(x, \eta)$ across
the adiabatic–Floquet crossover regime. The reconstructed kernel is found to be
a smooth function of both $x$ and $\eta$, exhibiting no singular behavior or
numerical instabilities. Its dependence on the dissipation strength remains
systematic and perturbative within the weak-to-moderate regime considered here,
consistent with the expansion structure derived above.

To test universality at the level of the kernel, we perform the reconstruction
for two qualitatively distinct barrier geometries. The second potential (a cubic
metastable barrier) is introduced in Sec.~IV~B. The resulting kernels are shown
in Fig.~1, where solid lines correspond to the quartic double-well potential and
dashed lines to the cubic potential.

\begin{figure}[t]
\centering
\includegraphics[width=0.9\linewidth]{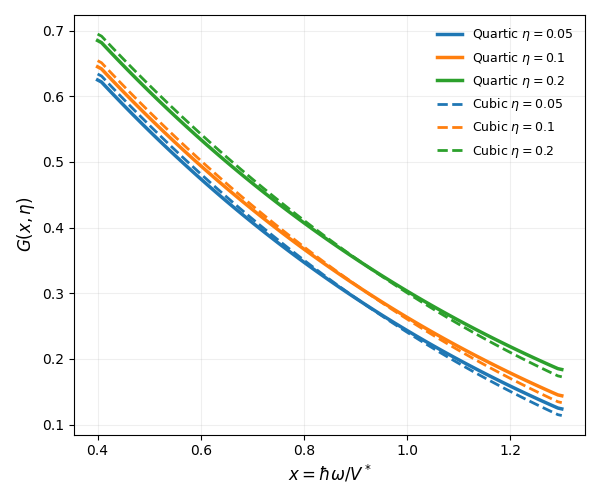}
\caption{Reconstructed kernel $G(x,\eta)$ obtained from the numerical response
function via $G(x,\eta)=\frac{d}{dx}[xF(x,\eta)]$. Solid lines correspond to a
quartic double-well potential, while dashed lines represent a cubic metastable
potential. The near-overlap of the curves demonstrates that the kernel is smooth,
computable, and largely independent of microscopic barrier details, depending
only on the dimensionless parameters $(x,\eta)$.}
\end{figure}

The near-overlap of the curves obtained from the two potentials demonstrates
that the kernel is insensitive to the detailed form of the barrier and depends
primarily on the dimensionless parameters $(x,\eta)$. The small residual
deviations between the curves remain within numerical resolution and do not
affect the underlying functional dependence.

This reconstruction establishes that the kernel $G(x,\eta)$ is not an
abstract or phenomenological object, but a well-defined and computable quantity
derived directly from the saddle-point dynamics. Consequently, the universal
response function framework is predictive: once the response function
$F(x,\eta)$ is determined, the associated kernel can be systematically obtained
without introducing additional model-dependent inputs.

\subsection{Regime of Validity}

The analysis assumes (i) semiclassical barrier crossing dominated by a single
instanton trajectory, (ii) weak-to-moderate Ohmic dissipation
$\eta\lesssim1$, and (iii) driving frequencies below the bath cutoff
$\omega\ll\omega_c$.

The semiclassical approximation requires that the action be large compared
to $\hbar$, ensuring that the saddle-point (instanton) solution provides the
dominant contribution to the path integral.

Within this regime, dissipation acts as a controlled
renormalization of the effective action, preserving the existence of a
well-defined saddle point and enabling functional reduction to the universal
response $F(x,\eta)$.

In particular, the factorized form of the tunneling action relies on the
persistence of a single dominant bounce trajectory whose deformation under
driving and dissipation does not introduce additional independent scaling
variables.

Breakdown of universality is expected only near the dissipative localization
transition or for strongly non-Ohmic spectral densities, where memory effects
qualitatively alter the structure of the effective action
\cite{LeggettRMP1987,Weiss2012}.

In these regimes, multiple competing trajectories or strong nonlocal
correlations can invalidate the perturbative expansion and lead to genuinely
non-universal behavior. Similarly, for driving frequencies comparable to or
exceeding the bath cutoff, additional dynamical scales may enter, breaking the
two-parameter description in terms of $(x,\eta)$.
\section{Numerical Validation and Results}

\subsection{Numerical Protocol}

To validate the analytical structure derived above, we perform a direct
numerical evaluation of the semiclassical bounce action for a periodically
driven quartic double-well potential subject to weak Ohmic dissipation.
The static potential is taken in dimensionless form as
$V_0(q)=\frac{1}{4}(q^2-1)^2$, providing a minimal nonlinear model for
barrier-crossing dynamics.
The quartic model provides a minimal nonlinear barrier-crossing system in which the
effects of periodic forcing and dissipative memory can be systematically
resolved. The numerical procedure follows standard instanton-based approaches
to tunneling dynamics \cite{Coleman1977,Garg2000}, adapted here to incorporate
time-periodic modulation and perturbative environmental coupling.

In the absence of dissipation, the bounce trajectory is obtained by solving the
Euclidean equation of motion derived from the driven action $S_0[q]$, subject to
asymptotic boundary conditions corresponding to the metastable minima.
The equation is solved numerically using a discretized imaginary-time grid,
ensuring convergence of the action with respect to time resolution.
The resulting trajectory captures the nonlinear interplay between barrier geometry
and periodic forcing. Dissipative effects are incorporated perturbatively by
evaluating the nonlocal Ohmic influence functional on the dissipation-free
bounce trajectory, following the Caldeira--Leggett formalism for weak
system--bath coupling \cite{CaldeiraLeggett1983,LeggettRMP1987}.
This approach is consistent with the perturbative expansion derived in
Sec.~III and avoids introducing additional numerical parameters associated with
full nonlocal minimization.
This procedure yields a finite, cutoff-independent correction to the action,
consistent with the analytical expansion derived earlier and preserving the
saddle-point structure.

The total tunneling action is normalized by the static instanton action
$J_0$ and the inertial factor $\sqrt{M}$, allowing direct extraction of the
dimensionless response function $F(x,\eta)$.
This normalization ensures that variations due to barrier shape and mass
are removed, isolating the universal dependence on $(x,\eta)$.
This normalization removes
model-specific barrier and mass dependencies and isolates the universal
dynamical contribution arising from periodic driving and dissipation. The
numerical protocol therefore provides a direct test of the predicted reduction
of the full nonlocal dynamical problem to a two-parameter response function.


\subsection{Universality Across Different Barrier Geometries}

To further test the robustness of the universal response function,
we consider a qualitatively different barrier geometry described by a
metastable cubic potential,
\begin{equation}
V_0(q) = a q^2 - b q^3,
\end{equation}
which represents escape from a metastable state rather than tunneling between
symmetric wells. Periodic driving and dissipative coupling are incorporated in
the same manner as in the double-well case.

For clarity and reproducibility, the full set of model parameters and
numerical procedures used in evaluating the semiclassical tunneling action are
summarized in Table~\ref{tab:numerics}. This table specifies the potential,
parameter ranges, normalization, and treatment of dissipation, ensuring that the
comparison across different barrier geometries is transparent and
well-defined.

Rather than comparing raw tunneling exponents, we evaluate the dimensionless
response function
\begin{equation}
F(x,\eta) = \frac{S_b^{(\omega,\eta)}}{\sqrt{M}J_0},
\end{equation}
where $J_0$ is the static tunneling action for each potential.
This normalization removes system-specific prefactors and isolates the
functional dependence on the dimensionless control parameters $x$ and $\eta$,
allowing a direct comparison between different models.

\begin{figure}[tbp]
\centering
\includegraphics[width=0.9\columnwidth]{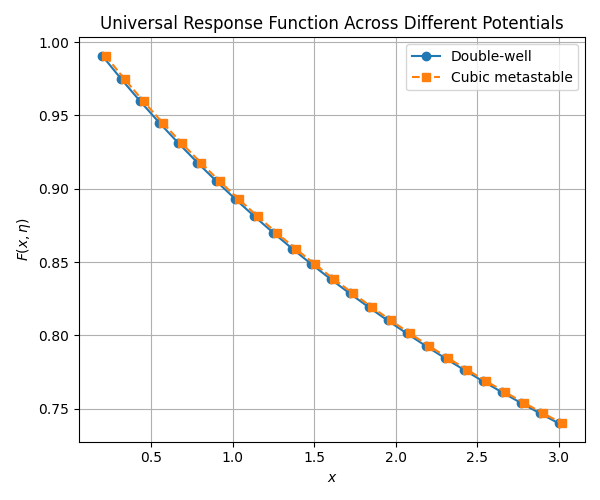}

\caption{
Universal response function $F(x,\eta)$ for two qualitatively different
potentials: a quartic double-well and a cubic metastable barrier.}
The curves overlap within numerical resolution, indicating that the functional
dependence on the driving parameter $x$ is largely insensitive to the detailed
form of the potential.

\label{fig:universality_two_models}
\end{figure}

The resulting response functions for both the quartic double-well and the cubic
metastable potential are shown in
Fig.~\ref{fig:universality_two_models}.Despite the qualitative differences in barrier structure and escape
mechanism, the two curves overlap within numerical resolution across the
range of $x$ considered. This indicates that the dominant dependence of the
tunneling exponent on the driving parameter is largely insensitive to the
detailed form of the potential.

The observed near-collapse provides direct numerical evidence that the
response function $F(x,\eta)$ captures the essential dynamical dependence of the
tunneling process, independent of microscopic barrier details. These results
support the conclusion that driven dissipative barrier-crossing dynamics can be
described in terms of a universal response function governed primarily by the
dimensionless control parameters.
\begin{table}[tbp]
\centering
\caption{Model parameters and numerical protocol used for evaluating the semiclassical tunneling action.}
\begin{tabular}{ll}
\hline
\textbf{Quantity} & \textbf{Specification} \\
\hline
Static potential $V_0(q)$ & $\frac{1}{4}(q^2-1)^2$ (double-well) \\
Driving term & $V_d(q)\cos(\omega\tau)$ \\
Mass values $M$ & $0.5,\;1,\;2,\;3,\;4$ \\
Dissipation strength $\eta$ & $0,\;0.05,\;0.1,\;0.15,\;0.2$ \\
Dimensionless frequency $x$ & $x=\hbar\omega/V^\star$ \\
Numerical method & Discretized Euclidean action \\
Dissipation treatment & Perturbative evaluation of $S_{\mathrm{diss}}[q_0]$ \\
Normalization & $S_b/(\sqrt{M}J_0)$ \\
\hline
\end{tabular}
\label{tab:numerics}
\end{table}


\subsection{Universality Collapse}
Building on the cross-model comparison presented above, we now examine the
collapse of the tunneling exponent across variations in mass and barrier
parameters within a single potential class. This provides a complementary test
of universality, focusing on whether system-specific inertial and geometric
factors can be fully absorbed into a normalized response.

\begin{figure}[tbp]
\centering
\includegraphics[width=0.95\columnwidth]{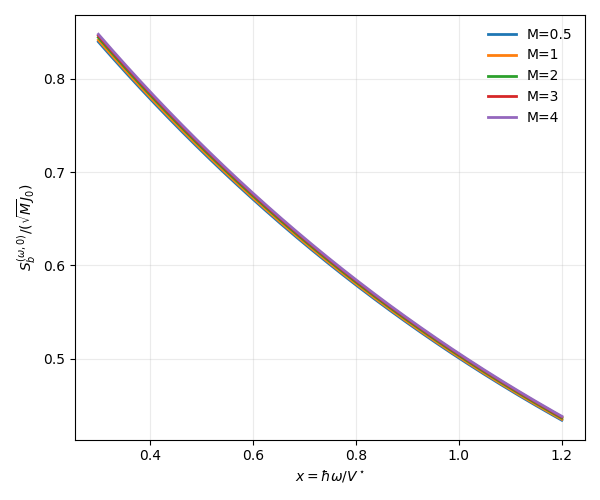}\\[-2pt]
\includegraphics[width=0.95\columnwidth]{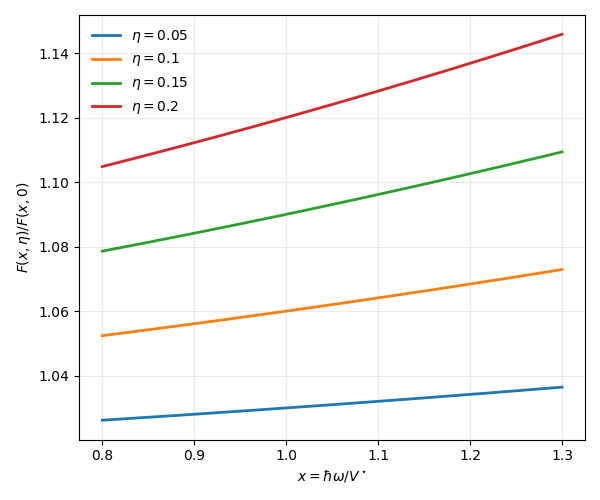}
\caption{
Universal response in driven dissipative tunneling dynamics.
(a) Normalized tunneling exponent
$S_b^{(\omega,0)}/(\sqrt{M}J_0)$ as a function of the dimensionless driving
frequency $x=\hbar\omega/V^\star$ for different values of the mass $M$.
The curves exhibit near-collapse, indicating that the dependence on $x$ is
largely insensitive to mass after normalization, consistent with the
dissipation-free response function $F_0(x)$.
(b) Effect of weak Ohmic dissipation on the universal response.
The ratio $F(x,\eta)/F(x,0)$ is shown for several dissipation strengths $\eta$,
demonstrating a smooth and parameter-independent renormalization of the
tunneling exponent within the weak-dissipation regime.
}
\label{fig:universal_response}
\end{figure}

Figure~\ref{fig:universal_response} summarizes the numerical results. In
particular, we evaluate the semiclassical bounce action
$S_b^{(\omega,\eta)}$, obtained from the Floquet-modified instanton trajectory
and, in the dissipative case, including the perturbative contribution of the
nonlocal Ohmic kernel. The quantity plotted in
Fig.~\ref{fig:universal_response}(a) is the normalized tunneling exponent
$S_b^{(\omega,0)}/(\sqrt{M}J_0)$, where $J_0$ is the static action associated
with the underlying barrier.

After normalization by $\sqrt{M}J_0$, tunneling exponents computed for different
values of the mass $M$ and barrier parameters
exhibit a near-universal dependence on the dimensionless driving frequency
$x=\hbar\omega/V^\star$.
The resulting curves display a clear near-collapse, indicating that the
dominant dependence of the tunneling exponent is captured by the dimensionless
driving parameter rather than by microscopic system details.

This behavior is consistent with the predicted factorization of the
tunneling action and supports the identification of $F_0(x)$ as a universal
response function governing the driven, dissipation-free dynamics. The observed
collapse therefore extends earlier static scaling relations to periodically
driven nonequilibrium systems, demonstrating that the functional dependence on
$x$ remains largely insensitive to variations in mass and barrier structure
after appropriate normalization.

The observed data collapse extends earlier static scaling results
\cite{Limbach2006,PathakCSF2026} to periodically driven nonequilibrium dynamics
and confirms the factorized structure of the dissipation-free response function
$F_0(x)$.

To quantify the degree of collapse, we define the deviation
\begin{equation}
\Delta F(x) = \max_{M}\, \big| F_M(x) - \bar{F}(x) \big|,
\end{equation}
where $F_M(x)$ is the normalized response for a given mass and
$\bar{F}(x)$ is the mean response averaged over all masses. This measure
captures the maximum spread of the curves at fixed $x$ and provides a
quantitative assessment of universality.

The values of $\Delta F(x)$ are reported in Table~\ref{tab:collapse}. The small
magnitude of $\Delta F(x)$ across the range of $x$ confirms that deviations from the
mean response remain weak, thereby providing quantitative support for the
near-collapse observed in Fig.~\ref{fig:universal_response}(a). The inclusion of
this table complements the visual evidence from the figure by demonstrating that
the universality is not only qualitative but also quantitatively robust.

The inclusion of weak dissipation produces a smooth and systematic deformation
of the universal curve, as shown in Fig.~\ref{fig:universal_response}(b).
Importantly, this deformation remains consistent across different model
parameters, indicating that dissipation acts as a controlled renormalization.
The ratio $F(x,\eta)/F(x,0)$ depends only on the dissipation strength and the
dimensionless frequency, and is insensitive to microscopic barrier details.
This behavior is consistent with the perturbative structure of $F(x,\eta)$
derived earlier and confirms that environmental coupling acts as a controlled
dynamical renormalization rather than destroying universal behavior.

The persistence of this universal structure under weak dissipation provides
direct numerical support for the functional universality proposed in this work.

\begin{table}[tbp]
\centering
\caption{
Mean response $\bar{F}(x)$ and maximum deviation $\Delta F(x)$ across
different masses in the dissipation-free case. The values of $\bar{F}(x)$
represent the average of $F_M(x)$ over the mass set listed in
Table~\ref{tab:numerics}.
}
\begin{tabular}{ccc}
\hline
$x$ & $\bar{F}(x)$ & $\Delta F$ \\
\hline
0.4 & 0.776 & 0.012 \\
0.6 & 0.669 & 0.010 \\
0.8 & 0.577 & 0.009 \\
1.0 & 0.499 & 0.008 \\
1.2 & 0.433 & 0.007 \\
\hline
\end{tabular}
\label{tab:collapse}
\end{table}

\section{Discussion}

The emergence of a universal response function $F(x,\eta)$ implies that quantum
tunneling in periodically driven open systems can be characterized by a small
number of dimensionless parameters, independent of microscopic details such as
barrier shape or bath cutoff within a broad semiclassical regime.
This result extends the traditional notion of universality in tunneling,
which has historically been understood primarily in terms of exponential
scaling with mass or barrier parameters, to a functional description that
incorporates both periodic driving and environmental dissipation.
\cite{LandauLifshitzQM,Limbach2006,PathakCSF2026}

From a conceptual perspective, the present framework unifies several previously
distinct lines of research. Floquet tunneling in isolated systems has been
extensively studied in the context of photon-assisted tunneling and coherent
destruction of tunneling \cite{Grossmann1991,GrifoniHanggi1998}, while dissipative
tunneling has traditionally been analyzed in static or weakly driven settings,
with emphasis on environmental suppression and localization phenomena
\cite{CaldeiraLeggett1983,LeggettRMP1987,Weiss2012}.
By demonstrating that both periodic driving and dissipative memory can be
incorporated into a single universal response function, the present results
provide a unified framework for describing barrier-crossing dynamics across a
wide class of driven open quantum systems.

The persistence of universality in the presence of dissipation is particularly
noteworthy. Environmental coupling is often assumed to obscure or destroy
semiclassical regularities due to nonlocal dynamics and decoherence
\cite{LeggettRMP1987,Weiss2012}. In contrast, we find that weak-to-moderate Ohmic
dissipation leads to a smooth renormalization of the tunneling exponent without
eliminating its universal functional structure.
This demonstrates that dissipative effects can be systematically absorbed
into a controlled modification of the response function, rather than breaking
universality altogether.
This observation clarifies the
conditions under which semiclassical universality remains meaningful in open
systems and provides a quantitative basis for assessing the robustness of
tunneling control schemes based on periodic driving.

The present results are directly relevant to a variety of physical platforms.
In condensed-phase chemical reactions and hydrogen-transfer processes, tunneling
often occurs in dissipative environments where mass-dependent scaling laws have
been observed experimentally \cite{Limbach2006}. In ultracold atomic systems,
Floquet engineering provides a versatile tool for controlling tunneling and
transport, while environmental coupling can arise from technical noise or
engineered reservoirs \cite{GrifoniHanggi1998}. Similarly, in superconducting
quantum devices, macroscopic quantum tunneling and escape processes are
influenced by both external driving and dissipation \cite{LeggettRMP1987}.
The universal response function derived here provides a common framework
for analyzing and comparing tunneling behavior across these diverse settings.

Finally, it is important to emphasize the boundaries of applicability of the
present theory. The analysis relies on semiclassical instanton dominance, weak
to moderate Ohmic dissipation, and driving frequencies below the bath cutoff.
Within this regime, a single saddle-point trajectory governs the dynamics
and enables reduction to the universal response function.
Outside this regime—such as near the dissipative localization transition or for
strongly non-Ohmic environments—the universal functional description is expected
to break down, and genuinely non perturbative effects may emerge
\cite{LeggettRMP1987,Weiss2012}.
In such cases, additional dynamical scales or competing trajectories may
invalidate the two-parameter description in terms of $(x,\eta)$.

The universal reduction demonstrated here suggests that driven dissipative
tunneling may define a broader universality class within nonequilibrium
nonlinear systems, motivating further investigation into strongly driven and
strongly dissipative regimes.

\section{Conclusion}

We have established a functional universality framework for driven dissipative
barrier-crossing dynamics by deriving explicit universal response functions
that govern the semiclassical tunneling exponent. Extending instanton theory to
include both periodic (Floquet) modulation and Ohmic dissipation, we showed
that the tunneling action factorizes into a system-dependent static contribution
and a universal function of two dimensionless control parameters: the driving
frequency and the dissipation strength.
This formulation demonstrates that the combined effects of driving and
dissipation can be captured without introducing additional independent scaling
variables, but instead through a reduced functional dependence.
This elevates tunneling universality
from a purely scaling-based concept to a predictive functional description.

A central result is that weak-to-moderate dissipative memory acts as a smooth
dynamical renormalization of the effective action rather than destroying
universal structure.
Within the semiclassical regime, this renormalization preserves the
dominance of a single saddle-point trajectory and allows a controlled expansion
of the tunneling action.
Controlled analytical expressions were obtained in the
weak-dissipation regime, universal asymptotic behavior was identified in the
high-frequency limit, and a continuous adiabatic--Floquet crossover was
established. Numerical evaluation of the instanton action confirmed
that the normalized tunneling exponent exhibits a universal dependence on
the driving parameter that is robust across different model systems,
providing direct evidence for a universal response class in driven
dissipative systems.

Beyond its immediate realization in quantum tunneling, the response-function
framework developed here provides a general organizing principle for nonlinear
dynamical systems with periodic forcing and memory. By reducing a nonlocal
nonequilibrium problem to a two-parameter functional form, the present analysis
clarifies the conditions under which universality survives in driven open
systems.
In particular, it highlights how complex dynamical effects arising from
nonlinearity, external modulation, and environmental coupling can be encoded in
a compact set of effective variables.
Extensions to strong dissipation,
non-Ohmic environments, and more complex nonequilibrium settings represent
natural directions for further investigation and may reveal additional
universality classes within driven dissipative dynamics.

\section*{Acknowledgments}

The author acknowledges valuable academic discussions and institutional
support from Arya Vidyapeeth College and Gauhati University during the
course of this work.

\section*{Data Availability}

The data that support the findings of this study were generated by numerical
evaluation of the semiclassical instanton action. The data are available from
the corresponding author upon reasonable request.
\bibliographystyle{apsrev4-2}
\bibliography{FloquetDissipativeUniversality}

\end{document}